\documentclass[prl,aps,letterpaper,floatfix,superscriptaddress,nofootinbib,preprintnumbers,twocolumn,10pt]{revtex4-1}

\usepackage{dcolumn}
\usepackage{bm}
\usepackage[dvips]{graphicx}
\usepackage{epsfig}
\usepackage{slashed}
\usepackage{feynmp}
\usepackage{hhline}
\usepackage{amsmath, amsfonts, amssymb}
\usepackage[normalem]{ulem}
\usepackage[breaklinks=true]{hyperref}
\hypersetup{
  colorlinks=true,
  linkcolor=blue,
  citecolor=blue,
  urlcolor=blue
}

\newcommand{\be}{\begin{equation}}
\newcommand{\ee}{\end{equation}}
\newcommand{\bee}{\begin{equation*}}
\newcommand{\eee}{\end{equation*}}
\newcommand{\bea}{\begin{eqnarray}}
\newcommand{\eea}{\end{eqnarray}}
\newcommand{\bean}{\begin{eqnarray*}}
\newcommand{\eean}{\end{eqnarray*}}

\usepackage{color}

\begin{document}


\begin{flushleft} 
IFT-UAM-CSIC-20-154 
\end{flushleft}
\begin{flushright} 
FTUAM-20-24
\end{flushright}

\vspace{-2mm}

\title{Shining light through the Higgs portal with $\gamma\gamma $ colliders}

\author{A. Garcia-Abenza} 
\affiliation{Instituto de F\'isica Fundamental, CSIC,
Serrano 121, 28006, Madrid, Spain}

\author{J. M. No} 
\affiliation{Instituto de F\'isica Te\'orica, IFT-UAM/CSIC,
Cantoblanco, 28049, Madrid, Spain}
\affiliation{Departamento de F\'isica Te\'orica, Universidad Aut\'onoma de Madrid,
Cantoblanco, 28049, Madrid, Spain}


\date{\today}

\begin{abstract}
High-energy $\gamma\gamma$ colliders constitute a potential running mode of future  $e^+ e^-$ colliders such as the ILC and CLIC. We study the sensitivity of a high-energy $\gamma\gamma$ collider to the Higgs portal scenario to a hidden sector above the invisible Higgs decay threshold. 
We show that such $\gamma\gamma$ collisions could allow to probe the existence of dark sectors through the Higgs portal comparatively more precisely than any other planned collider facility, from the unique combination of sizable cross-section with clean final state and collider environment. In addition, this search could cover the singlet Higgs portal parameter space yielding a first-order electroweak phase transition in the early Universe.
\end{abstract}

\maketitle




\noindent \textbf{I. Introduction.} The existence of dark sectors in Nature, uncharged under the gauge symmetries of the Standard Model (SM), and interacting with the SM through the Higgs boson $h$ is a well-motivated possibility: both theoretically, since the operator $H^{\dagger}H$ is the only super-renormalizable SM Lorentz invariant operator singlet under the SM gauge symmetries~\cite{Patt:2006fw}, and in connection to open problems in particle physics and cosmology, like the nature of dark matter (DM)~\cite{Bertone:2004pz,Silveira:1985rk,McDonald:1993ex,Burgess:2000yq}. In addition, a singlet scalar field $S$ coupled to the SM via the Higgs portal Lagrangian interaction $\left|H\right|^2 S^2$ (with $H$ the SM Higgs doublet) is arguably the simplest possible extension of the SM, further motivated by the fact that it could yield a strongly first-order electroweak (EW) phase transition in the early Universe~\cite{Profumo:2007wc,Espinosa:2011ax,Curtin:2014jma,Chen:2017qcz}, possibly allowing for EW baryogenesis as the origin of the observed matter-antimatter asymmetry of the Universe~\cite{Espinosa:2011eu,Cline:2012hg} (see~\cite{Morrissey:2012db} for a review).  

\begin{figure}[h]
\begin{center}
\includegraphics[width=0.21\textwidth]{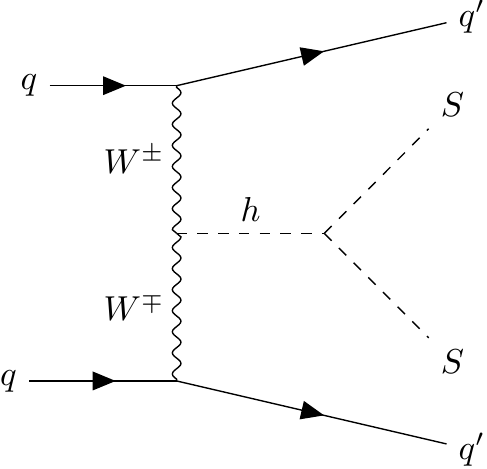}
\hspace{2mm}
\includegraphics[width=0.21\textwidth]{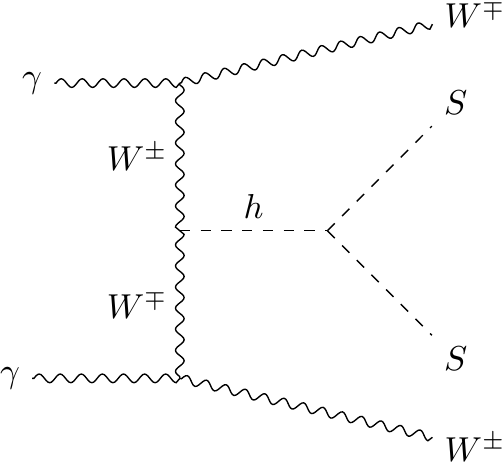}
\caption{\small Feynman diagrams for singlet scalar $S$ pair production through the Higgs portal. Left: hadron colliders (for $e^+ e^-$ colliders, initial state fermions would be $e^{\pm}$, and final state fermions would be neutrinos). Right:  $\gamma\gamma$ colliders.}
\label{Feyn_Diagrams}
\end{center}

\vspace{-5mm}

\end{figure}

Despite its simplicity and appeal, such a singlet Higgs portal scenario is very challenging to probe experimentally\footnote{If $S$ is not itself the DM particle, 
since otherwise direct detection DM constraints on the singlet scalar scenario apply~\cite{Cline:2013gha}.} at high-energy colliders when the Higgs boson decay $h \to S S$ is not kinematically open ({\sl i.e.}~for singlet scalar masses $m_s$ above the decay 
threshold). 
At the Large Hadron Collider (LHC) it is possible to directly probe the hidden sector in final states with hadronic jets and missing transverse energy $\slashed{E}_T$, via the vector-boson-fusion (VBF) process $p p \to j j\, + S S$~\cite{Craig:2014lda,Curtin:2014jma} (see Fig.~\ref{Feyn_Diagrams}-left), with the pair of singlet scalars giving rise to $\slashed{E}_T$. The High-Luminosity LHC would however only be sensitive to very large values of the Higgs portal coupling~\cite{Craig:2014lda,Curtin:2014jma}.  
A future {\sc FCC-}hh~\cite{Benedikt:2018csr} hadron collider operating at a center-of-mass (c.o.m.) energy $\sqrt{s} = $ 100 TeV would improve on the LHC sensitivity~\cite{Craig:2014lda,Golling:2016gvc}, profiting from the large enhancement of the VBF off-shell Higgs process at high energy. Yet, the messy hadronic environment hinders very strong sensitivity improvements.   

Future high-energy $e^+ e^-$ colliders like the $\sqrt{s} = 1$ TeV International Lineal Collider (ILC) or the $\sqrt{s} = 1.5/3$ TeV Compact Linear Collider (CLIC) could provide the ideal setup to probe the Higgs portal above the $m_h/2$ threshold due to a combination of reach in energy and clean collision environment. However, for $e^+ e^-$ collisions the dominant VBF process to produce a pair of singlet scalars is $e^+ e^- \to \nu\nu S S$, thus completely invisible and impossible to trigger on at colliders.

In this letter, we show that a $\gamma \gamma$ (or $e^{\pm} \gamma$) operating mode of a high-energy $e^{+} e^-$ collider like ILC or CLIC would overcome the above problems, providing an optimal setup to probe the Higgs portal to a dark sector, via the process $\gamma \gamma \to W^+ W^- + \slashed{E}_T$, see Fig.~\ref{Feyn_Diagrams}-right. After discussing the key aspects of $\gamma \gamma$ colliders in section II, we introduce the singlet scalar extension of the SM in section III as the benchmark scenario for our study,
and briefly discuss its impact on the EW phase transition. We then analyze the sensitivity of an ILC and CLIC-based $\gamma\gamma$ collider to the Higgs portal above threshold scenario in section IV. 
\vspace{-2mm}

\noindent \textbf{II. $\gamma \gamma$ colliders.}  The possibility of a high-energy $\gamma\gamma$ (or $\gamma e$) collider based on a linear $e^{+}e^{-}$ collider has been considered since the early 1980's~\cite{Ginzburg:1981ik,Ginzburg:1981vm,Ginzburg:1982yr}. The physical principle is the generation of high-energy photons through Compton back-scattering of laser photons by the high-energy electrons or positrons of the $e^{+}e^{-}$ collider beams, a mechanism that has subsequently been extensively studied (see e.g.~\cite{Telnov:1995hc,Telnov:1997vg,Telnov:1998vs,Telnov:2000kq,Telnov:2000zx,Burkhardt:2002vh,Badelek:2001xb,Battaglia:2004mw,Yu:2018omk}).
In the conversion region a photon with energy $E_{0}$ 
is scattered on an electron with energy $E_{e}$ at a small collision angle $\alpha$ (almost head-on). The photons from Compton back-scattering have
a spectrum with maximum energy $E^{\mathrm{max}}_{\gamma}$ given by
\begin{equation}
E^{\mathrm{max}}_{\gamma} =\frac{\kappa}{1+\kappa}E_{e}\quad , \quad 
\kappa=\frac{4\,E_{e}\,E_{0}\cos^{2}\left(\alpha/2\right)}{m_{e}^{2}}
\label{eq:max_photon_energy}
\end{equation}
\noindent where $m_e$ is the electron mass. According to Eq.~\eqref{eq:max_photon_energy}, the largest possible laser frequency $\omega_0 = E_0/\hbar$ should be used in order to increase $E^{\mathrm{max}}_{\gamma}$. This also increases the fraction of hard photons in the spectrum~\cite{Telnov:1995hc}. However, at large $\kappa$ the resulting high-energy photons are then converted to $e^{+}e^{-}$ pairs in collisions with laser photons, so the optimum value $\kappa = \kappa^{\mathrm{max}}$ is the threshold of this conversion process, given for a head-on collision by~\cite{Telnov:1995hc,Yu:2018omk}
$E^{\mathrm{max}}_{\gamma} E_{0} = m_e^2$. Combining this threshold condition with 
Eq.~\eqref{eq:max_photon_energy} yields $\kappa^{\mathrm{max}}=2\, (1+\sqrt{2}) \approx 4.83$, resulting in a highest energy $E_{\gamma}^{\mathrm{max}} \approx 0.83 E_{e}$.

\begin{figure}[t]
\begin{centering}
\includegraphics[width=0.48\textwidth]{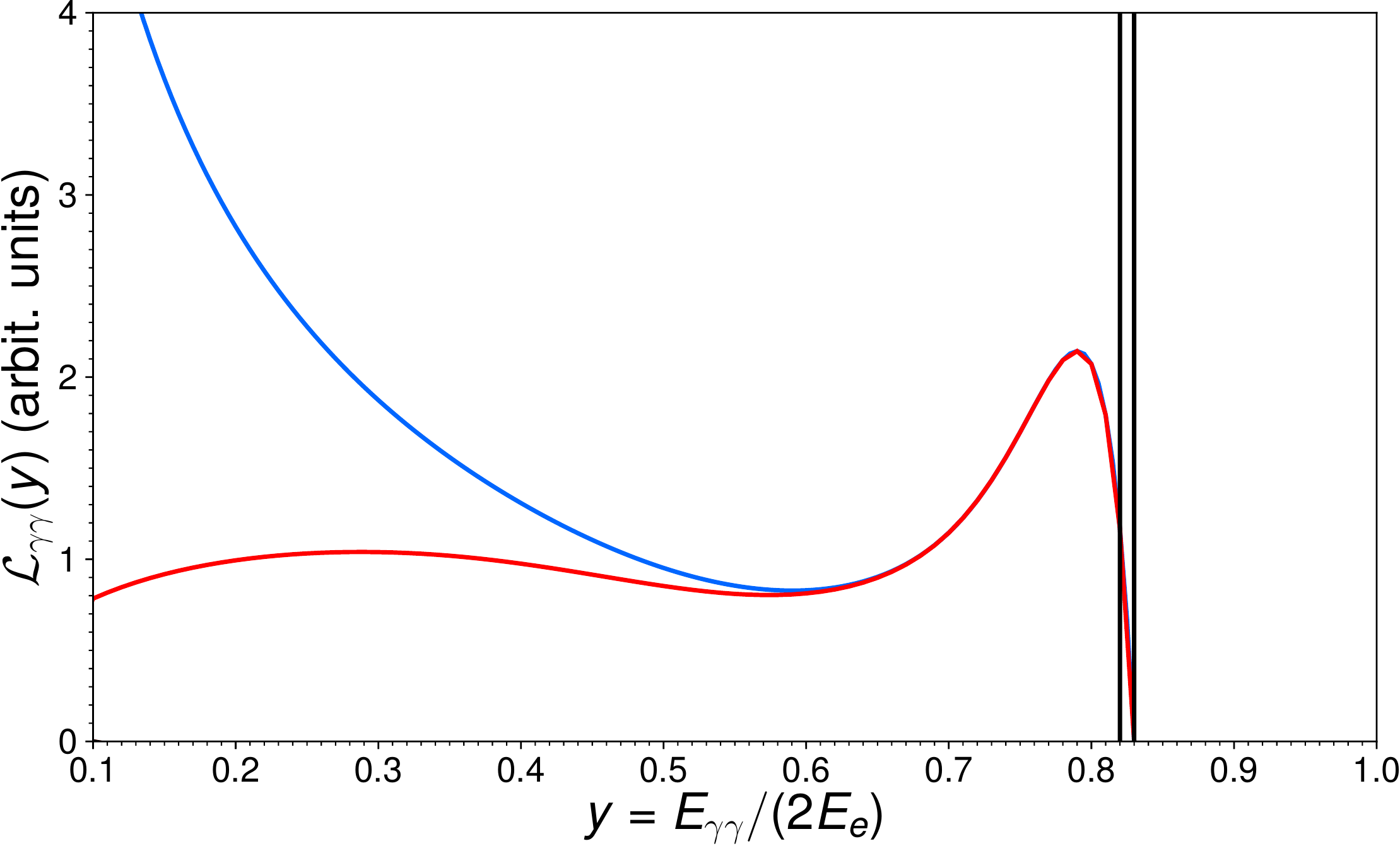} 

\vspace{1mm}

\includegraphics[width=0.48\textwidth]{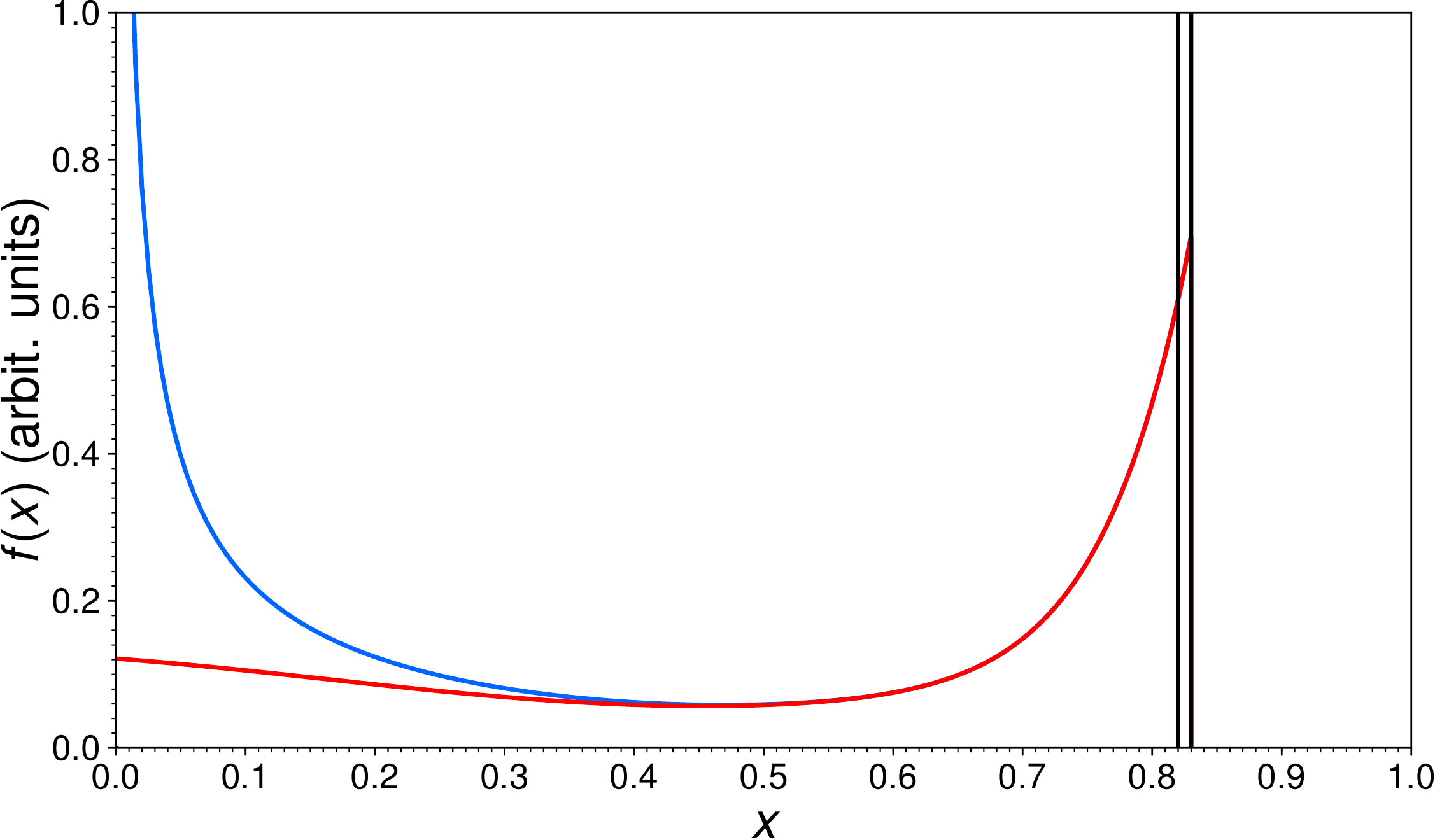} 

\vspace{-2mm}

\caption{TOP: Luminosity spectra $\mathcal{L}_{\gamma\gamma}(y)$ used in this work: idealized spectrum $\propto \delta(y - y^{\rm max})$ (black); analytic spectrum $\mathcal{L}_{\gamma\gamma}^{nc} \left(y\right)$ for $\kappa^{\mathrm{max}} = 4.8$ and $2\lambda_{e}P_{\gamma}=-1$ (red); spectrum $\mathcal{L}_{\gamma\gamma}^{c} \left(y\right)$ including multiple Compton scattering effects and beamstrahlung (blue). All of then are normalized to the same value of $\mathcal{L}_{\gamma\gamma}\left(y \,> 0.8 \,y^{\rm max}\right)$ (see text for details).
BOTTOM: Photon energy distribution $f(x)$ for each of the above luminosity spectra $\mathcal{L}_{\gamma\gamma}(y)$.}
\label{fig2:spectra}
\par\end{centering}

\vspace{-4mm}

\end{figure}

The energy spectrum of the resulting photon beam is peaked at $E_{\gamma}^{\mathrm{max}}$, and the number of high energy photons dramatically increases for polarised beams with $2\lambda_{e}P_{\gamma}=-1$, being $\lambda_{e}$ ($|\lambda_{e}| \leq 1/2$) the mean helicity of the initial electron and $P_{\gamma}$ that of the laser photon.
In addition to this high-energy peak there is also a factor 5-8 larger (in luminosity) low-energy spectrum which is produced by multiple Compton scattering and beamstrahlung photons. These low-energy collisions have a large longitudinal
boost in the detector reference frame. 
The $\gamma\gamma$ luminosity $\mathcal{L}_{\gamma\gamma}$ 
in the high-energy part of the spectrum is proportional to the geometric
luminosity of the electron beams $\mathcal{L}_{G}$~\cite{Telnov:1998vs,Badelek:2001xb}. Considering $ y = E_{\gamma\gamma}/(2E_{e})$, one approximately has~\cite{Telnov:1998vs,Telnov:1997vg} 
\begin{equation}
\mathcal{L}_{\gamma\gamma}\left(y \,> 0.8 \,y^{\rm max}\right)
\approx 0.1\, \mathcal{L}_G \, ,
\label{lumi_aa_hard}
\end{equation}
where the maximum possible value of $y$ is given by $y^{\rm max}  = E_{\gamma}^{\mathrm{max}}/E_{e} \simeq 0.83$.
As discussed in~\cite{Telnov:1997vg,Burkhardt:2002vh,Battaglia:2004mw}, luminosities $\mathcal{L}_{\gamma\gamma}
\left(y \,> 0.8 \,y^{\rm max}\right) \sim 10^{34}$ cm$^{-2}$ s$^{-1}$ (and perhaps up to $10^{35}$ cm$^{-2}$ s$^{-1}$~\cite{Telnov:1997vg}) could be reached at a multi-TeV $\gamma \gamma$ collider, comparable to those of the HL-LHC.

In the rest of this work we consider three different 
$\mathcal{L}_{\gamma\gamma}\left(y\right)$
spectra for a multi-TeV $\gamma\gamma$ collider:
{\sl (i)} An idealized spectrum, with the energy of the back-scattered photons essentially localized at $E_{\gamma}^{\mathrm{max}} = 0.83 E_{e}$, {\sl i.e.}~$\mathcal{L}_{\gamma\gamma}\left(y\right) \propto \delta(y - y^{\rm max})$.~{\sl (ii)} An analytic high-energy $\gamma\gamma$ collider luminosity spectrum for $\kappa^{\mathrm{max}} = 4.8$ and $2\lambda_{e}P_{\gamma}=-1$ without multiple Compton scattering and beamstrahlung effects
obtained from~\cite{Badelek:2001xb} and labelled here $\mathcal{L}_{\gamma\gamma}^{nc} \left(y\right)$.~{\sl (iii)}:~A 
$\sqrt{s} = 3$ TeV luminosity spectrum including the effect at low energies from multiple Compton scatterings and beamstrahlung  
fitted from~\cite{Burkhardt:2002vh}, labelled here $\mathcal{L}_{\gamma\gamma}^{c} \left(y\right)$.
The three spectra are shown jointly in Fig.~\ref{fig2:spectra} (top).
From these luminosity spectra we derive the respective photon energy distributions $f(x)$ (see Fig.~\ref{fig2:spectra} (bottom)), given by 
\be
\mathcal{L}_{\gamma\gamma}\left(y\right) = \int_{0}^{x^{\mathrm{max}}} \hspace{-5mm} dx_1  \int_{0}^{x^{\mathrm{max}}} \hspace{-5mm} dx_2 \, f(x_1) \,f(x_2) \,\, \delta(y - \sqrt{x_1x_2})
\ee
with $x^{\rm max} = y^{\rm max}$. The luminosity spectra and photon energy distributions 
will then be used together with a Monte Carlo event generation from {\sc MadGraph 5}~\cite{Alwall:2014hca} and {\sc Whizard}~\cite{Kilian:2007gr,Kilian:2018onl} to construct event samples for the SM background and our BSM signal in section IV.

%
%







\vspace{2mm}

\noindent \textbf{III. The singlet scalar extension of the SM.} The simplest realization of the Higgs portal to a dark sector consists of an extension of the SM by a real scalar singlet field $S$ \cite{Silveira:1985rk, McDonald:1993ex,Profumo:2007wc, Espinosa:2011eu} which is odd under a $\mathbb{Z}_{2}$ symmetry. 
The scalar potential for the theory is
\begin{eqnarray}
V\left(H,S\right) = &-&\mu_{H}^{2}|H|^{2} + \lambda_{H} |H|^{4}  + \frac{\mu_{S}^{2}}{2}\,S^{2} +\frac{\lambda_{S}}{4}S^{4} \nonumber \\ &+& \lambda_{HS}\,|H|^{2}S^{2}\,,
\label{eq:VHS}
\end{eqnarray}
with $H = (0, (v+ h)/\sqrt{2})$ and $v=246$~GeV the EW scale. 
After EW symmetry breaking, the $\mathbb{Z}_{2}$ symmetry is preserved  
for $m_S^2 = \mu_{S}^{2} + \lambda_{HS} \, v^2  > 0$.
In this case, the singlet scalar does not mix with the SM Higgs boson after EW symmetry breaking and only interacts with the SM through its portal coupling $\lambda_{HS}$ to the Higgs boson. In particular, if $m_S > m_h/2 \simeq 63$ GeV, the $h \to S S$ Higgs boson decay into two singlet scalars is forbidden and the only way to access the hidden sector directly (to produce $S$) is via an off-shell Higgs~\cite{Craig:2014lda,Curtin:2014jma}, which makes this scenario very challenging to probe at colliders.
\vspace{1mm}

As outlined in the introduction, extending the SM by the singlet scalar field $S$ may impact the breaking of EW symmetry in the early Universe:~in the SM the EW phase transition is found to be a smooth-cross over process using non-perturbative methods~\cite{Kajantie:1996mn,Csikor:1998eu}; it would then not induce the needed departure from thermal equilibrium to generate the matter-antimatter asymmetry at the EW scale. The presence of the singlet field $S$ together with a sizable portal coupling $\lambda_{HS}$ may dramatically change this conclusion,   
triggering a first-order EW phase transition strong enough to allow for baryogenesis~\cite{Profumo:2007wc,Espinosa:2011ax,Curtin:2014jma,Espinosa:2011eu} or produce a stochastic background of gravitational waves observable by LISA (see~\cite{Caprini:2015zlo,Caprini:2019egz} and references therein). The combined Higgs-singlet field dynamics in the early Universe may yield a first-order EW phase transition already through the interplay of tree-level and thermal effects, via a two-step symmetry breaking process~\cite{Espinosa:2011ax,Patel:2012pi}: the $\mathbb{Z}_2$ symmetry would be broken first along the $S$ field direction and restored later, when EW symmetry breaking occurred. The evolution of the potential minimum
$(\left\langle S \right\rangle, \left\langle H \right\rangle)$ from high to low temperature would be $(0, 0) \to (0,\, w_T) \to (v_T,\,0)$, with $v_T$ and $w_T$ respectively the Higgs and singlet vevs at finite temperature $T$. The potential barrier between $\left(0,w_T\right)$ and $\left(v_T,0\right)$ minima would induce a strongly first-order phase transition.
%
%
The lowest value of $\lambda_{HS}$ as a function of $m_S$ for which such a two-step first-order EW phase transition occurs has been obtained in~\cite{Curtin:2016urg} including 1-loop corrections and higher-order thermal effects (which qualitatively preserve the above picture). Such $\lambda_{HS}$ value provides a specific sensitivity target for future colliders~\cite{Ramsey-Musolf:2019lsf}. 

\vspace{2mm}

\begin{figure}[h]
\begin{centering}
\includegraphics[width=0.483\textwidth]{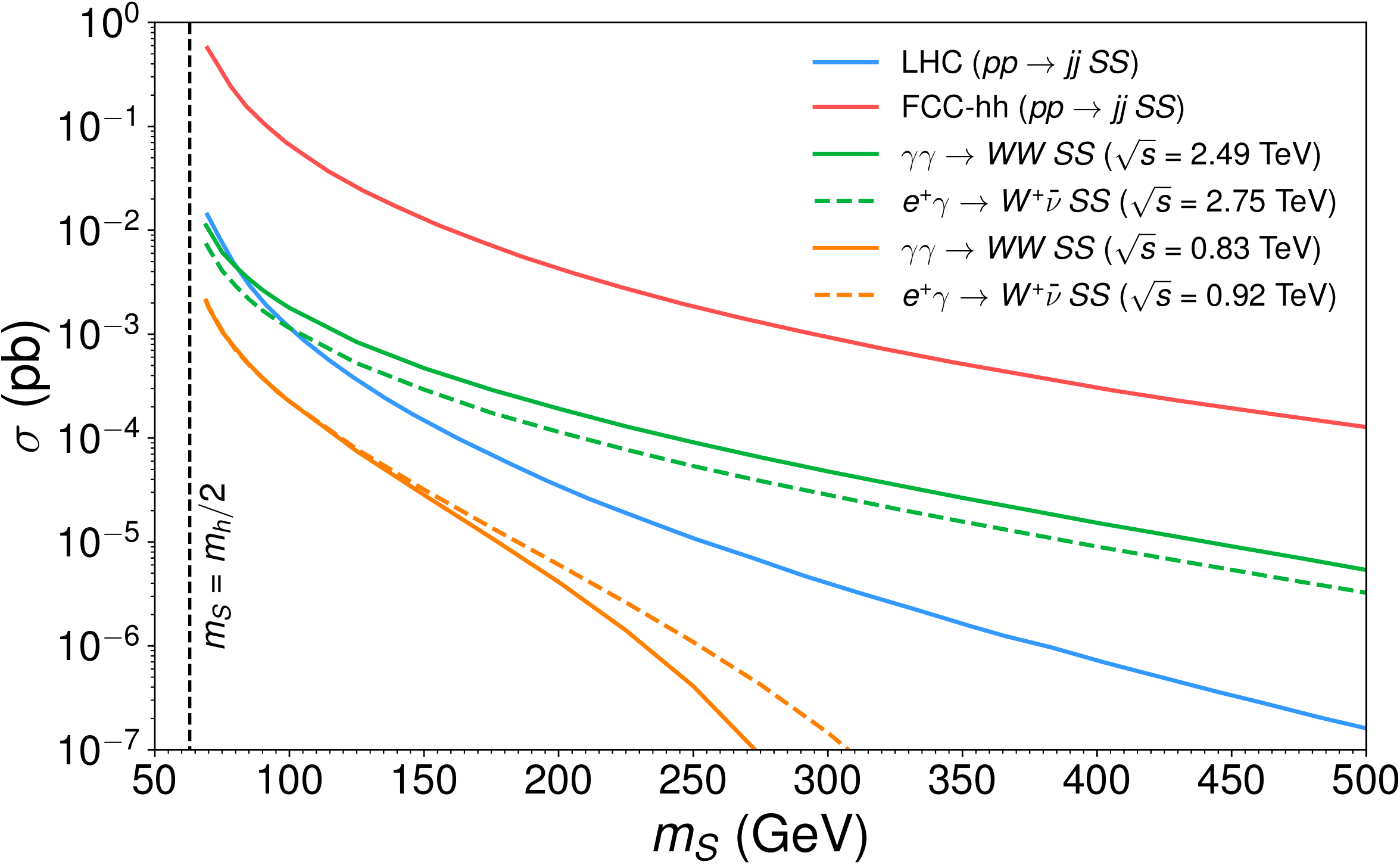} 
\caption{LO cross section for singlet pair production (for $\lambda_{HS}  = 1$) as a function of $m_S$ via $\gamma\gamma \to W^+ W- S S$ (solid) and $e^+\gamma \to W^+ \bar{\nu} S S$ (dashed) for a 3 TeV CLIC-based collider (green) and a 1 TeV ILC-based collider (orange). The $E_{\gamma}^{\mathrm{max}}/E_{e} = 0.83$ factor is explicitly taken into account. Also shown are the LO cross sections for 14 TeV LHC (blue) and 100 TeV FCC-hh (red) via the process $p p \to j j S S$.}
\label{XS_colliders_Singlet_mass}
\par\end{centering}

\vspace{-4mm}

\end{figure}

\noindent \textbf{IV. Collider Analysis.} We now investigate the sensitivity that a $\gamma\gamma$ collider based on ILC or CLIC could achieve in probing the Higgs portal scenario to a dark sector discussed in the previous section above the kinematic decay threshold for $h \to S S$ (that is, for $m_S > 63$ GeV), via the process 
$\gamma\gamma \to W^+ W^- S S$.   
First, we show in Fig.~\ref{XS_colliders_Singlet_mass} the (pair) production cross section of the singlet scalar $S$ at a $\sqrt{s} = 1$ TeV ILC and a $\sqrt{s} = 3$ TeV CLIC (including the $\sqrt{s}|_{\gamma\gamma}/\sqrt{s}|_{ee} = E_{\gamma}^{\mathrm{max}}/E_{e} =  0.83$ reduction factor for $\gamma\gamma$ collisions) for the idealized luminosity spectrum from Fig.~\ref{fig2:spectra}, as a function of the singlet scalar mass $ m_S$ and for $\lambda_{HS} = 1$. We also show the corresponding $\gamma e$ production cross sections via the process $e^{+}\gamma  \to W^{+} \bar{\nu} S S$, and include for comparison the respective production cross sections for $\sqrt{s} = 14$ TeV LHC and a future FCC-hh at $\sqrt{s} = 100$ TeV via the process $p p \to j j \,S S$. All cross sections are obtained at leading order (LO). For a 3 TeV CLIC-based $\gamma\gamma$ collider in particular, the cross section becomes much larger than that of LHC as $m_S$ increases, and is $\sim 50$ times smaller than that of FCC-hh, yet the signal cross section ratio to the SM background is much more favorable than in the latter, due to the cleaner environment of a lepton/$\gamma$ collider as compared to a hadron collider. 

\vspace{-2mm}

Considering hadronic decays for the $W$-bosons, the dominant SM background to the $\gamma\gamma \to W^+ W^- S S$ signal comes from triboson production $\gamma\gamma \to W^+ W^- Z$ with $Z \to \nu\bar{\nu}$.  
Both for the $1$ TeV ILC and $3$ TeV CLIC analyses, we generate our signal and SM background samples at LO with {\sc MadGraph 5}~\cite{Alwall:2014hca}, requiring parton-level jets to satisfy $p_T^j > 20$ GeV and $\left|\eta_j \right| < 4.5$. Generation is done for the three $\gamma\gamma$ luminosity spectra from Fig.~\ref{fig2:spectra}. For the non-idealized spectra, we perform a fine discretization of $\mathcal{L}_{\gamma\gamma}(y)$, generate event samples for $\sqrt{s} = y$ and appropriately re-weight and combine the various samples to include the effect of the photon energy distributions $f(x)$ in the $\gamma\gamma$ collisions (for each event, we fix the longitudinal boost in the laboratory frame via a random generation according to $f(x)$ and the corresponding $y$).   

For the extraction of the signal, we introduce the ``missing invariant mass" $m_{\rm miss}$:
\begin{equation}
m_{\rm miss}^2 = \left(\sqrt{\hat{s} + (p_{z_{WW}} +\slashed{p}_z)^2} - E_{WW}\right)^2 - \slashed{E}_T^2 - \slashed{p}_z^2 \,,
\end{equation}
with $\sqrt{\hat{s}}$ the c.o.m.~energy of the partonic collision, $E_{WW} = \sqrt{m_W^2 + \left|\vec{p}_{W^+} \right|^2 } + \sqrt{m_W^2 + \left|\vec{p}_{W^-} \right|^2 } $ the energy of the $W^+W^-$ system, $p_{z_{WW}} = p_{z_{W^+}} + p_{z_{W^-}}$ the sum of longitudinal momenta of the $W$-bosons and $\slashed{p}_z$ the longitudinal component of the missing momentum. Both $E_{WW}$ and $p_{z_{WW}}$ can be accurately reconstructed from the hadronic $W$ decay products.
For the idealized luminosity spectrum $\mathcal{L}_{\gamma\gamma}(y) \propto \delta(y - y^{\rm max})$, knowledge of the $\gamma\gamma$ collision c.o.m. energy together with the condition $\slashed{p}_z = - p_{z_{WW}}$ (absence of longitudinal boost for the collisions in the laboratory frame) allow to very efficiently disentangle the signal from the SM background: the reconstructed events peak around $m_{\rm miss} = m_Z$ for the SM background while having a lower bound $m_{\rm miss} \geq 2\, m_S$ for the signal. A 
cut $m_{\rm miss} > 160$ GeV suppresses the SM background below the $\mathcal{O}(1)\%$ level, while retaining a large signal fraction for $m_S > 63$ GeV. The corresponding $2\sigma$ exclusion sensitivity $S/\sqrt{B} = 2$ (with $S$ and $B$ the respective number of signal and background events) for $\lambda_{HS}$ as function of $m_S$ is shown in Fig.~\ref{figfinal:bounds}.   

\vspace{-2mm}

\begin{figure}[h]
\begin{centering}
\includegraphics[width=0.45\textwidth]{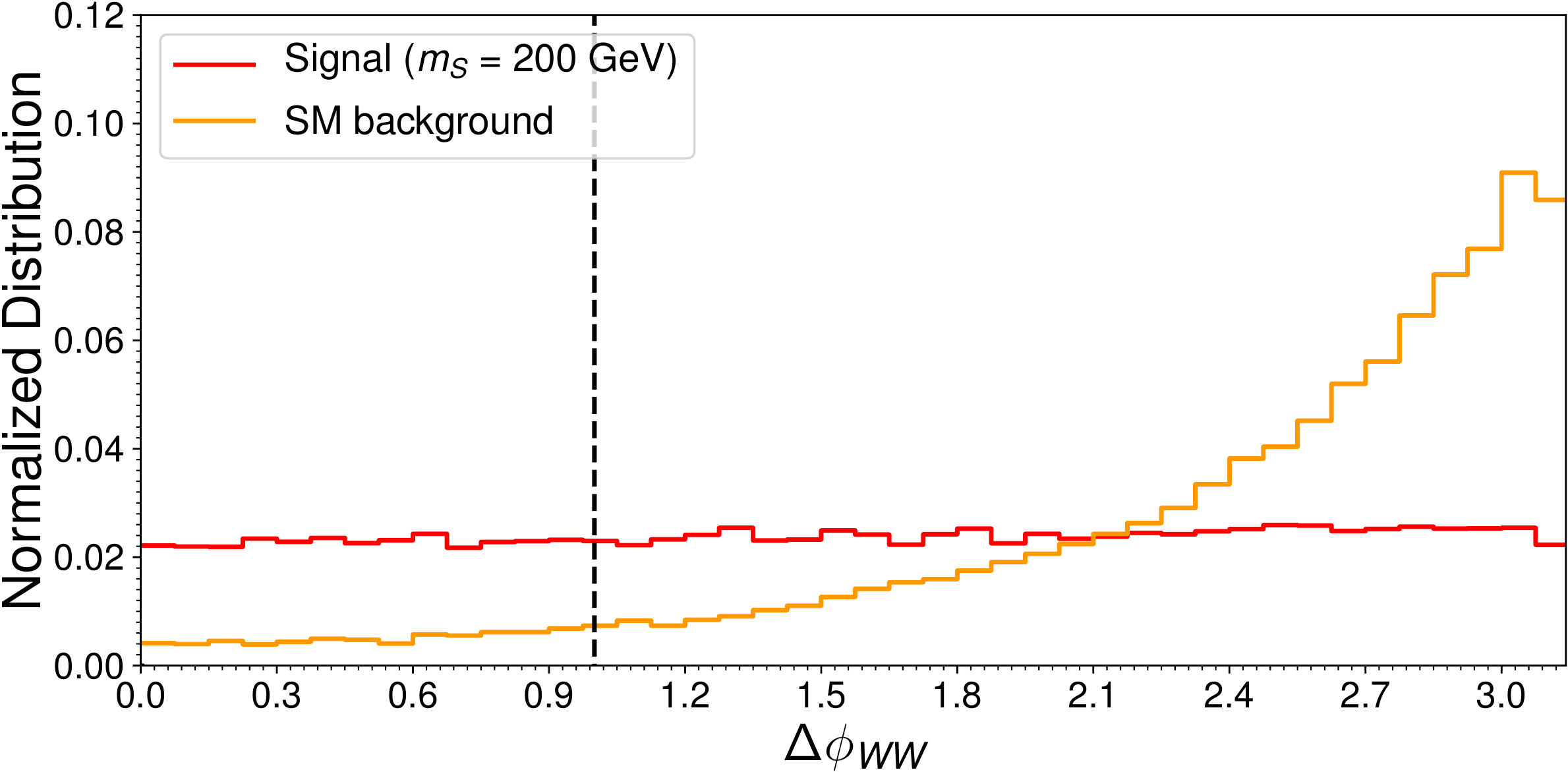}

\vspace{1mm}

\includegraphics[width=0.235\textwidth]{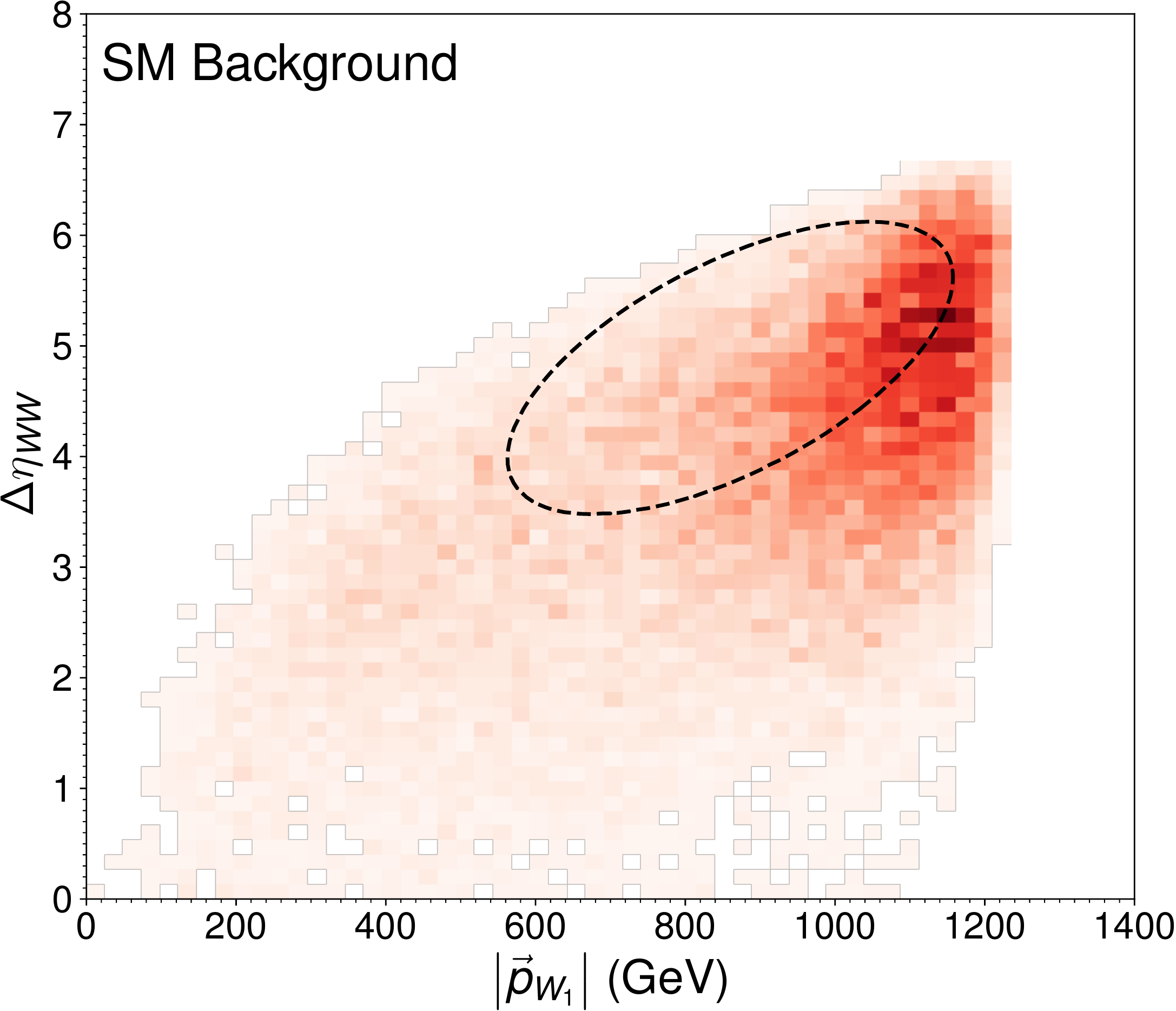}
\includegraphics[width=0.235\textwidth]{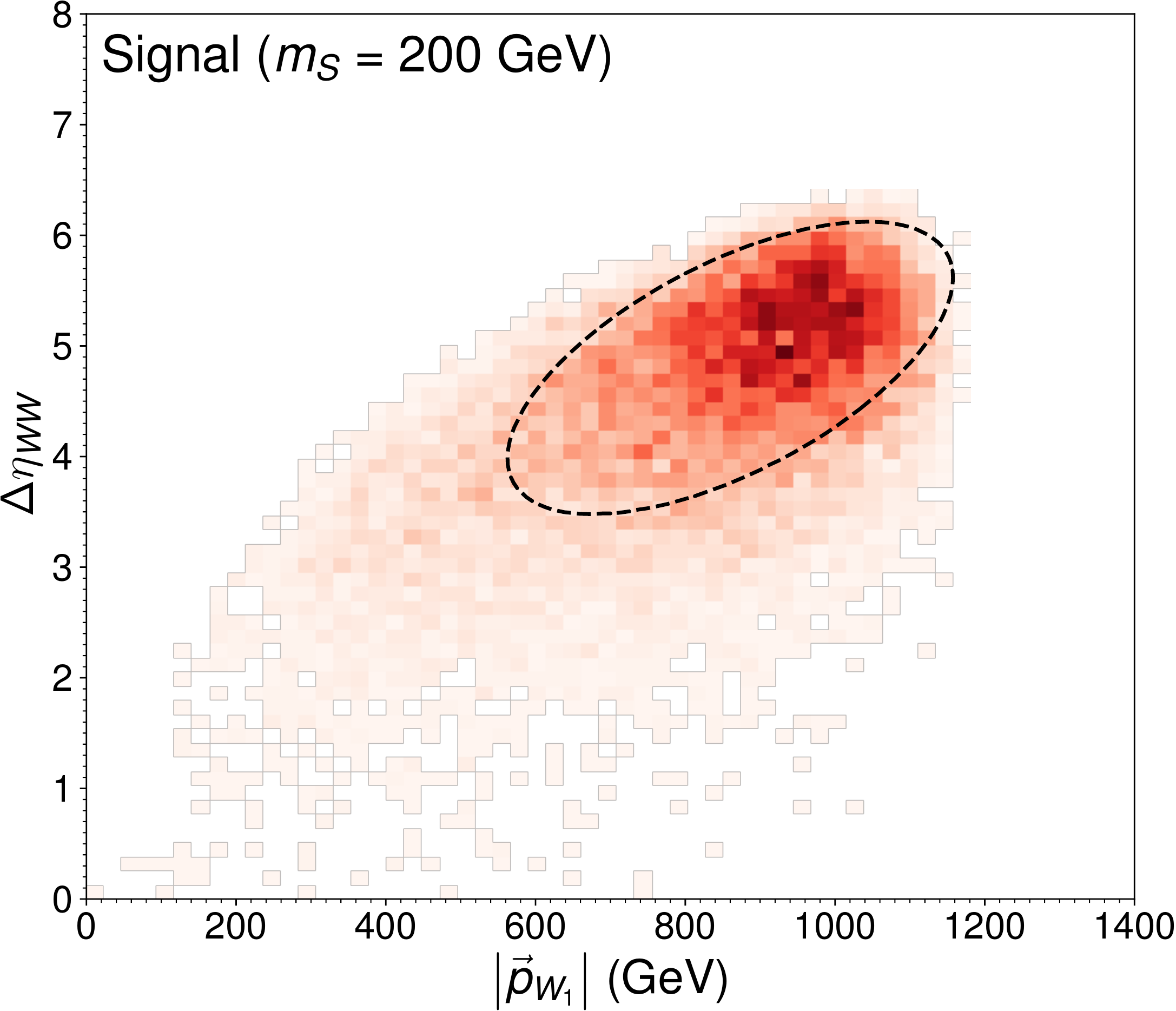} 

\vspace{1mm}

\includegraphics[width=0.235\textwidth]{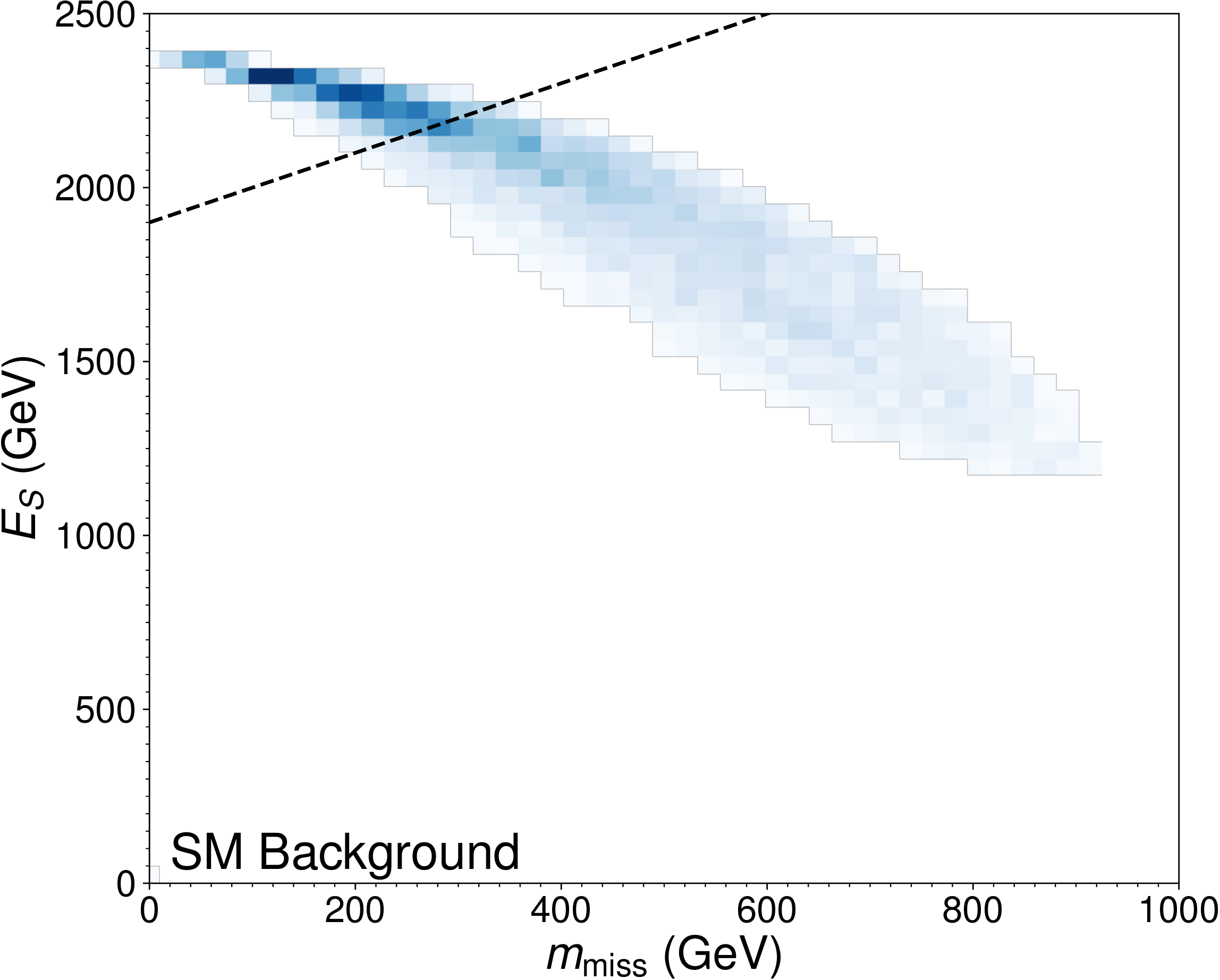}
\includegraphics[width=0.235\textwidth]{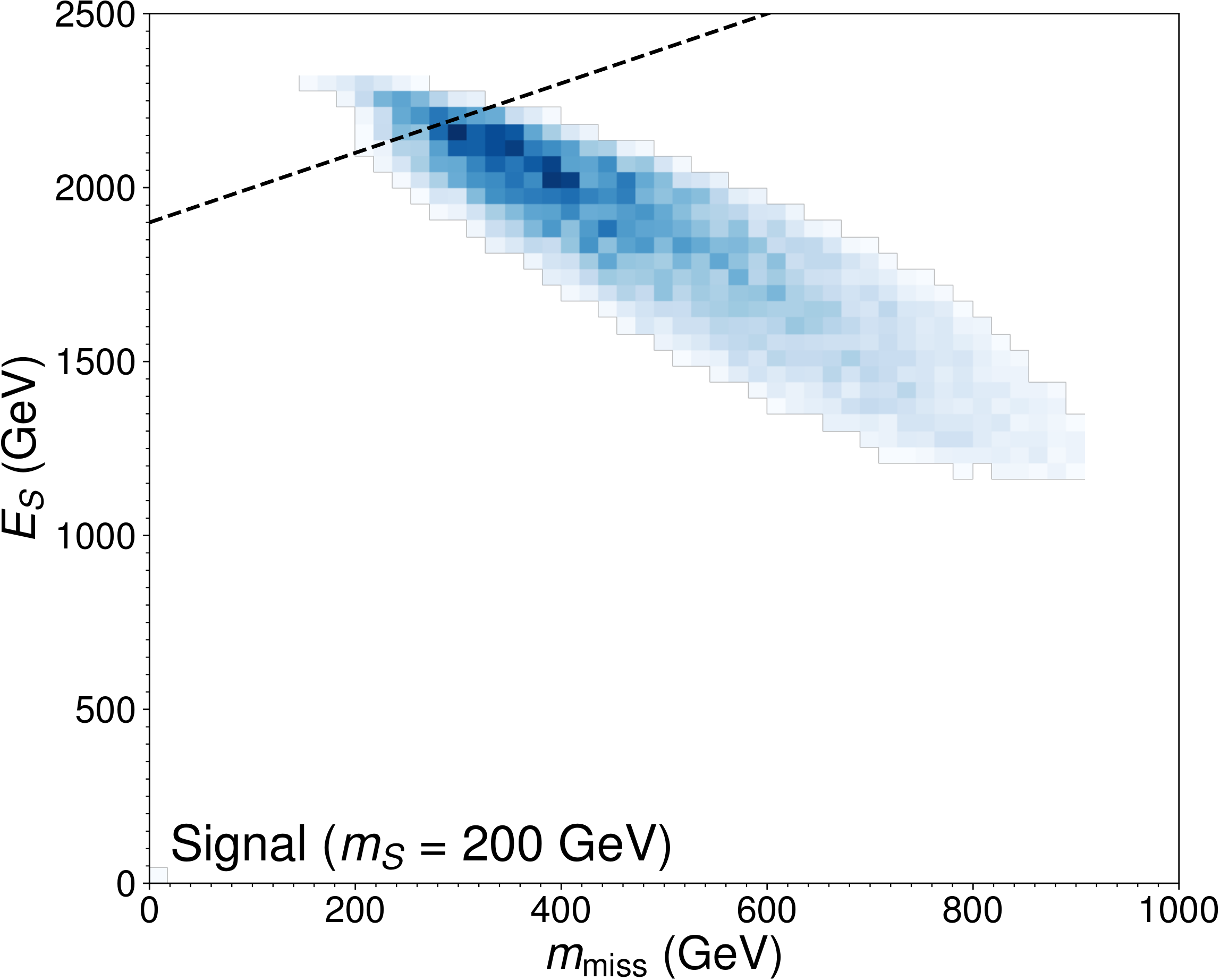} 

\vspace{-3mm}

\caption{$\mathcal{L}_{\gamma\gamma}^{nc} (y)$, $\sqrt{s}|_{ee} = 3$ TeV events. Top: Normalized $\Delta \phi_{WW}$ distribution for the SM background (orange) and $m_S = 200$ GeV signal (red). The selection cut $\Delta \phi_{WW} < 1$ is also shown (dashed-black line). Middle: $\left| \vec{p}_{W_1}\right|$ vs $\Delta \eta_{WW}$ distribution for the SM background (left) and $m_S = 200$ GeV signal (right) after $\Delta \phi_{WW}$ selection. Signal selection is shown as a dashed-black ellipse (see text for details). Bottom: $m_{\rm miss}$ vs $E_S$ distribution for the SM background (left) and $m_S = 200$ GeV signal (right) prior to the final signal region selection $m_{\rm miss} > E_S - E_0$ (depicted as a dashed-black line).}
\label{fig:Distros_SM}
\par\end{centering}

\vspace{-4mm}

\end{figure}

For the non-idealized luminosity spectra, $\sqrt{\hat{s}}$ is not known, and neither is the longitudinal boost of each collision in the laboratory frame. Yet, the above strategy is still useful, but needs to be preceded by an event selection to increase the signal significance, since the $m_{\rm miss}$ reconstruction is degraded in this case. Focusing on $\mathcal{L}_{\gamma\gamma}^{nc} (y)$ and $\sqrt{s}|_{ee} = 3$ TeV for concreteness
we show in Fig.~\ref{fig:Distros_SM} (top) the angular separation of the two hadronic $W$s in the transverse plane $\Delta \phi_{WW}$, for the SM background  and $m_s = 200$ GeV signal. The clear difference between signal and background is due to the different spin nature of the intermediate particle ($h$ vs $Z$) and we select events with $\Delta \phi_{WW} < 1$. After this selection, we show in  Fig.~\ref{fig:Distros_SM} (middle) the momentum of the hardest $W$-boson $\left| \vec{p}_{W_1}\right|$ vs the rapidity difference between $W$s, $\Delta \eta_{WW}$. For the signal (right) the two variables are heavily correlated, and we require $(c_{\theta} X + s_{\theta} Y)^2/r_1^2 + (s_{\theta} X - c_{\theta} Y)^2/r_2^2 < 1$, with $\theta = 0.3$, $r_1 = 3.1$, $r_2 = 1$, $X = \left| \vec{p}_{W}\right| / (100\,{\rm GeV})- c_1(m_S)$, $Y = \Delta \eta_{WW} - c_2(m_S)$. The functions $c_{1,2}(m_S)$ are fitted to the signal data, yielding: $c_1 = 9.8 - 0.41 \,m_S - 0.097 \,m_S^2$, $c_2 = 5.3 - 0.175 \,m_S - 0.042\, m_S^2$ ($m_S$ in units of $100$ GeV).  

Finally, we carry out the $m_{\rm miss}$ reconstruction for the surviving events.~We first remark that approximating $\sqrt{\hat{s}}$ purely via global event kinematic variables, e.g.~$\sqrt{\hat{s}} \sim \sqrt{s}_{{\rm min}}$~\cite{Konar:2008ei,Konar:2010ma} or $\sqrt{\hat{s}} \sim E_S \equiv E_{WW} + (\slashed{E}_T^2 + p_{z_{WW}}^2)^{1/2}$ (note that $E_S > \sqrt{s}_{{\rm min}}$), does not yield an acceptable $m_{\rm miss}$ reconstruction for both signal and SM background: the average difference $\sqrt{\hat{s}} - E_S$ is significantly larger for the signal than for the SM background, and this effect increases as $m_S$ grows. We use an averaged 
approximation $\sqrt{\hat{s}} \sim ( E_{\mathcal{L}} + \left| \vec{p}_{W_1}\right| + \left| \vec{p}_{W_2}\right| + (\slashed{E}_T^2 + p_{z_{WW}}^2)^{1/2})/2$, with $ E_{\mathcal{L}} = 2370$ GeV corresponding to the maximum of the $\mathcal{L}^{nc}_{\gamma\gamma} (y)$ spectrum (see Fig~\ref{fig2:spectra}-top). 
Assuming also $\slashed{p}_z \simeq - p_{z_{WW}}$, we show in Fig.~\ref{fig:Distros_SM} (bottom) the resulting distribution of $m_{\rm miss}$ vs $E_S$.  
Fig.~\ref{fig:Distros_SM} (bottom) highlights the degrading in the reconstruction of $m_{\rm miss}$ for non-idealized luminosity spectra, from the impossibility of accurately accessing $\sqrt{\hat{s}}$ and $\slashed{p}_z$ for each $\gamma\gamma$ collision. Still, defining the signal region as $m_{\rm miss} > E_S - E_0$ (see Fig.~\ref{fig:Distros_SM}), with fitted $E_0(m_S)/{\rm TeV} = 2.24 - 0.117 \, m_S - 0.028\, m_S^2$ ($m_S$ in units of $100$ GeV), improves the signal discrimination, particularly for large $m_S$.

The above analysis is repeated 
for the non-idealized luminosity spectrum $\mathcal{L}_{\gamma\gamma}^{c} (y)$.
In each case, we compute the $2\sigma$ exclusion sensitivity $S/\sqrt{B} = 2$ for $\lambda_{HS}$ as a function of $m_S$. These sensitivities are then shown in Fig.~\ref{figfinal:bounds}. The integrated luminosity we quote in each non-idealized scenario corresponds to that of the high-energy part of the $\gamma\gamma$ spectrum, $\mathcal{L}_{\gamma\gamma}\left(y \,> 0.8 \,y^{\rm max}\right)$ (recall the discussion around Eq.~\eqref{lumi_aa_hard}). We also show the $2\sigma$ exclusion sensitivities achievable at HL-LHC and FCC-hh via $p p \to j j + \slashed{E}_T$ obtained respectively from~\cite{Craig:2014lda} and~\cite{Curtin:2014jma}, as a comparison. In addition we depict in  Fig.~\ref{figfinal:bounds} the lowest value of $\lambda_{HS}$ compatible with a (two-step) first order EW phase transition~\cite{Curtin:2016urg}
in this scenario.~Fig.~\ref{figfinal:bounds} highlights that for comparable integrated luminosities, multi-TeV $\gamma\gamma$ collisions would directly probe dark sectors via the Higgs portal with precision similar to, and potentially higher than, future hadron colliders. A $\sqrt{s} = 3$ TeV - based $\gamma\gamma$ collider could cover the whole parameter space region compatible with a two-step singlet-driven strongly first-order EW phase transition that could allow for baryogenesis.




\vspace{1mm}

Before concluding, we stress the possibility of further enhancing the sensitivity to Higgs portal scenarios in $\gamma\gamma$ collisions by analyzing $W^{+} W^-$ semi-leptonic and/or leptonic final states. The contribution to the $\slashed{E}_T$ of the events from the $W$ decays in this case however demands a different strategy to suppress SM backgrounds (e.g. the use of transverse mass variables $M_T$ and $M_{T2}$~\cite{Lester:1999tx,Barr:2003rg}), and we leave such a study for the future.   

\begin{figure}[t]
\begin{centering}
\includegraphics[width=0.483\textwidth]{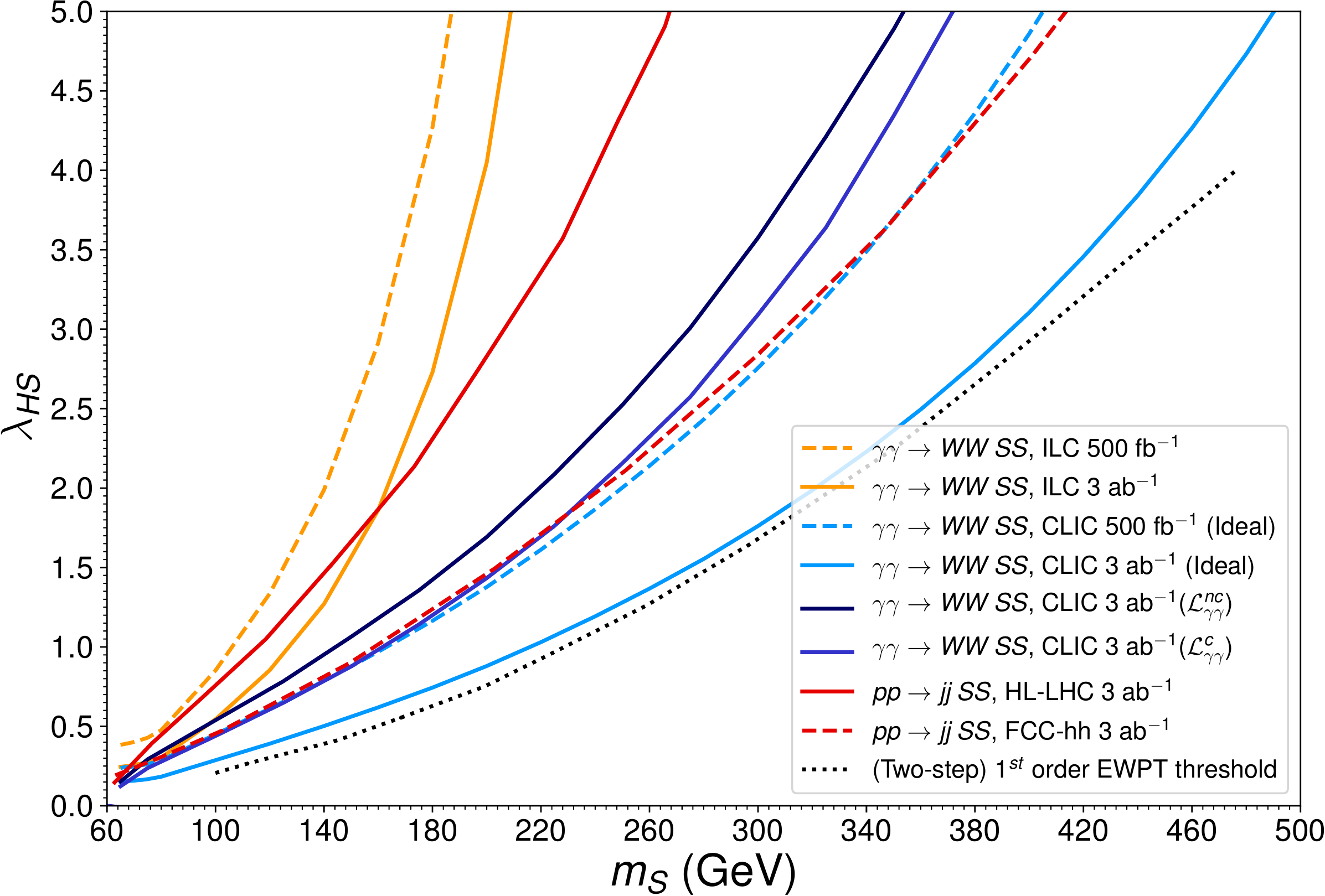}

\vspace{-4mm}

\caption{$2\sigma$ sensitivity in the ($\lambda_{HS}$, $m_S$) plane of the singlet Higgs portal model, for a $\gamma\gamma$ collider based on an $e^+ e^-$ c.o.m.~energy $\sqrt{s} = 1$ TeV  ("ILC") and $\sqrt{s} = 3$ TeV ("CLIC"). Different curves correspond to: ILC 500 fb$^{-1}$ with ideal $\mathcal{L}_{\gamma\gamma}$ (dashed-yellow), ILC 3 ab$^{-1}$ with ideal $\mathcal{L}_{\gamma\gamma}$ (solid-yellow), CLIC 500 fb$^{-1}$ with ideal $\mathcal{L}_{\gamma\gamma}$ (dashed-light-blue), CLIC 3 ab$^{-1}$ with ideal $\mathcal{L}_{\gamma\gamma}$ (solid-light-blue), CLIC 3 ab$^{-1}$ with $\mathcal{L}^{nc}_{\gamma\gamma}$ (solid-black) and CLIC 3 ab$^{-1}$ with  $\mathcal{L}^{c}_{\gamma\gamma}$  (solid-dark-blue). We also show the $2\sigma$ sensitivity of HL-LHC (solid-red) and FCC-hh (dashed-red) via the process $p p \to j j \,S S$. The dotted-black line shows the lower threshold for a two-step 1$^{st}$ order EW phase transition, as obtained from~\cite{Curtin:2016urg}.}
\label{figfinal:bounds}
\par\end{centering}

\vspace{-4mm}

\end{figure}





\vspace{-1mm}

\begin{center}
\textbf{Acknowledgements} 
\end{center}

\vspace{-1mm}

\begin{acknowledgments}
Feynman diagrams were drawn using {\sc TikZ-Feynman}~\cite{Ellis:2016jkw}.
J. M. N. was supported by Ram\'on y Cajal Fellowship contract RYC-2017-22986, and also acknowledges support from the Spanish MINECO's ``Centro de Excelencia Severo Ochoa" Programme under grant SEV-2016-0597, from the European Union's Horizon 2020 research and innovation programme under the Marie Sklodowska-Curie grant agreement 860881 (ITN HIDDeN) and from the Spanish Proyectos de I$+$D de Generaci\'on de Conocimiento via grant PGC2018-096646-A-I00. A. G.-A. thanks the MICIU,
for his grant within the ``Garant\'ia Juvenil'' program.

\end{acknowledgments}

\bibliography{GammaGamma.bib}{}

\end{document}